\documentclass[pra,twocolumn,letterpaper,showpacs]{revtex4}

\usepackage{graphics}
\usepackage{epsfig}

\begin{document} 

\title{Vortex line in a neutral finite-temperature superfluid Fermi gas}

\author{N.~Nygaard$^{1,2}$, G.~M.~Bruun$^3$, B.~I.~Schneider$^4$,
C.~W.~Clark$^1$, and D.~L.~Feder$^5$} 
\affiliation{$^1$Electron and Optical Physics Division, National
Institute of Standards and  Technology, Gaithersburg, Maryland
20899-8410}
\affiliation{$^2$Chemical Physics Program, University of Maryland,
College Park, Maryland} 
\affiliation{$^3$Niels Bohr Institute, Blegdamsvej 17, 2100
Copenhagen, Denmark} 
\affiliation{$^4$Physics Division, National Science Foundation, Arlington,
Virginia 22230}
\affiliation{$^5$Department of Physics and Astronomy, University of Calgary, 
Calgary, Alberta, Canada T2N 1N4}

\date{\today}

\begin{abstract} 

The structure of an isolated vortex in a dilute two-component neutral
superfluid Fermi  gas is studied within the context of self-consistent
Bogoliubov-de Gennes theory.  
Various thermodynamic properties are calculated and the shift in the
critical temperature due to the  
presence of the vortex is analyzed. The gapless excitations inside 
the vortex core are studied and a scheme to detect these states and
thus the presence of the vortex 
is examined.  The numerical results are 
compared with various analytical expressions when appropriate.

\end{abstract}


\maketitle

\section{Introduction}
The achievement of Fermi degeneracy in a confined gas of alkali 
atoms~\cite{DeMarco1999a,Truscott2001,Shreck2001,Granade2002,Hadzibabic2002,Roati2002,Jochim2002}   
has spurred great interest both theoretically and experimentally in
cold atomic gases with Fermi statistics.  
The atomic interactions are well-understood and often may be tailored
through the physics of Feshbach resonances by the application of
external magnetic 
fields~\cite{Feshbach1958,Feshbach1962a,Tiesinga1993a}. When
the atom-atom interaction  is attractive, the ground state of a two 
component gas is predicted to be superfluid at low
temperatures~\cite{Stoof1996a}. Such a  
superfluid would provide a unique test bed for the study and 
interpretation of analogous but much more complex systems, such as
superfluid $^3$He, unconventional superconductors, and neutron stars. 

One important  issue facing the cold atom community has been how
one would go about actually detecting the presence of superfluidity in  
these systems. Superfluidity in Bose-Einstein condensates (BECs) can
be inferred either by probing directly the momentum distribution of the
cloud, the collective modes (where the spectrum is strongly shifted
relative to the normal phase), or by generating quantized 
vortices (an unambiguous signature of the breakdown of irrotational 
flow) and simply viewing the associated ``holes'' in the particle  
density~\cite{Matthews1999a,Madison2000a}. Likewise, for superfluid
Fermi gases, the presence of superfluidity has been shown to give many
observable effects on the mode spectrum of the
gas~\cite{Baranov2000,Bruun2001c}.   
For fermions in the weak-coupling limit, the presence of a vortex would be very difficult to image
directly by looking at the density profile, as there is very little
depletion of the density in the vortex core~\cite{Nygaard2003}. However,  
the quantization of angular momentum which is a striking macroscopic 
effect of superfluidity can, as for bosons, be measured through the
energy shift of the quadrupole modes~\cite{Bruun2001a}.

Experimental techniques currently limit the temperature of trapped
Fermi gases to not much less than one tenth of the Fermi degeneracy
temperature $T_F$. The superfluid transition temperature $T_c$ of a
conventional uniform Bardeen, Cooper, Schrieffer (BCS) superconductor,
however, is typically lower: 
$T_c/T_F\simeq0.28e^{-\pi/2k_F|a|}\ll 1$, with  
$k_F$ the momentum at the Fermi surface, $a$ the $s$-wave scattering
length for low-energy two-body collisions, and $k_F|a|\ll 1$ in the
weak-coupling approximation where BCS theory is valid. A number of
schemes to raise $T_c$ to a value closer to temperatures already
accessible with dilute Fermi gases have recently been proposed. One of
these, referred to in the literature as ``resonance superfluidity''
involves tuning the scattering length to an extremely large value at a
Feshbach resonance~\cite{Timmermans2001,Holland2001}; recent experimental
results~(see for example \cite{Ohara2002,Bourdel2003,Strecker2003})  
show significant progress using this
approach, culminating in the production of a Bose-Einstein condensate of molcules~\cite{Greiner2003,Jochim2003b,Zwierlein2003b}. Another proposal involves loading the cold Fermi gas into a
three-dimensional optical lattice~\cite{Hofstetter2002}: if the
lattice is made sufficiently 
deep, the lowest-lying band will flatten to the point where all of the
atoms participate in the pairing, as opposed to regular BCS theory,
where only the small fraction of particles close to the Fermi surface
are available for pairing. Of course,
the lattice depth cannot be so great that coherence across 
the sample is destroyed, as has been observed for bosons in optical 
lattices~\cite{Anderson1998a,Orzel2001,Greiner2002}.  The inability to 
experimentally attain very low temperatures in dilute gases is
probably not fundamental, however. With an eye on future experiments,
it seems reasonable to explore the predictions of a
weak-coupling theory of Fermi superfluidity.

In the present manuscript, we examine in detail several properties of
the vortex phase of a neutral Fermi liquid using a microscopic weak
coupling theory. The theoretical framework is briefly discussed in Section~\ref{theory}, 
and we  present in Section~\ref{numerics} the details of our numerical procedure.  
Section~\ref{thermodynamics} is devoted to the calculation various thermodynamic 
quantities of the vortex phase, which are compared with the corresponding quantities in both the normal state and the superfluid with no vortex. Furthermore, we demonstrate that the vortex causes a shift of the superfluid transition temperature. Finally, in Section~\ref{laserprobe} we propose a way of observing the vortex through
``laser probing'' of the quasi-particle states trapped inside the
vortex core.

\section{Theoretical background}
\label{theory}
We consider a two component Fermi gas consisting of particles with
internal quantum numbers $\sigma=\uparrow, \downarrow$ 
and mass $m_a$ confined in a cylinder of length $L$ and radius $R$. 
For atomic
gases at low temperatures and realistic densities, the interactions far from
Feshbach resonances are characterized by the low energy parameter $a$ which 
is the $s$-wave scattering length appropriate for the scattering between the 
two specific internal states of the atoms. Therefore, only Fermi particles
in different  
internal states are able to interact. In our calculations, we assume an 
equal population of the two components
$N_\uparrow=N_\downarrow$ so that their densities $n_\sigma$ are
equal. The superfluid phase of the gas for $a<0$ can be described  
within mean field theory by the Bogoliubov-de Gennes (BdG)
equations~\cite{deGennes} 
\begin{eqnarray}
\left[\begin{array}{cc}{\mathcal{H}}^{\mathrm{HF}}-\mu&\Delta({\mathbf{r}})\\
\Delta^*({\mathbf{r}})&-({\mathcal{H}}^{\mathrm{HF}}-\mu) 
\end{array}
\right]\left[\begin{array}{c}u_\eta({\mathbf{r}})\\v_\eta({\mathbf{r}})\end{array}
\right]=
E_\eta\left[\begin{array}{c}u_\eta({\mathbf{r}})\\v_\eta({\mathbf{r}})\end{array}
\right].
\label{BdGeqs}
\end{eqnarray}
Here
${\mathcal{H}}^{\mathrm{HF}}=-\frac{\hbar^2}{2m_a}\nabla^2+V({\mathbf r})+gn_\sigma({\bf r})$ with the low energy effective coupling constant given by
$g=4\pi\hbar^2a/m_a$. The particle density and pairing field are defined as  
$n_{\sigma}({\mathbf{r}})=\langle\psi^{\dag}_{\sigma}({\mathbf r})
\psi^{\vphantom{\dag}}_{\sigma}({\mathbf r}) \rangle$ and
$\Delta({\mathbf{r}})=-g\langle\psi_{\uparrow}({\mathbf r})
\psi_{\downarrow}({\mathbf r}) \rangle$, respectively, where
$\psi_{\sigma}^{\dagger}({\mathbf r})$ is the usual fermionic field
operator creating a particle in the internal state $\sigma$ at
position ${\mathbf r}$. The Bogoliubov wave functions
$u_\eta({\mathbf{r}})$ and $v_\eta({\mathbf{r}})$ describe
quasi-particle excitations with energy $E_\eta>0$. The ultraviolet
divergence 
in the definition of the superfluid gap is regularized using the
pseudopotential method~\cite{Bruun1999a}. Since the system is essentially 
homogeneous, the spectrum is continuous, so we can use a
semi-classical version of this scheme also described
in~\cite{Bulgac2001a}. We augment 
it to incorporate the effect of the Hartree mean-field $gn_{\sigma}$ in order 
to achieve faster convergence of the solution. The same method was used
by Grasso and Urban~\cite{Grasso2003}, who present a detailed analysis
of the convergence properties.

\subsection{Vortex phase}
The superfluid order parameter is a complex number and can thus be
written as a real amplitude times a phase
\begin{equation}
\Delta({\mathbf r})=|\Delta({\mathbf r})|e^{i\theta({\mathbf r})}.
\end{equation} 
The superfluid velocity is then given by the spatial variation of the
phase of the order parameter
\begin{equation}
{\mathbf v}_s=\frac{\hbar}{2m_a}\nabla\theta({\mathbf r}),
\end{equation}
where $2m_a$ is the mass of a Cooper pair. 
For rotational currents,
the order parameter must vanish at the center of rotation where the superfluid
velocity diverges. Far away from the core, the flow velocity decreases with 
the distance $\rho$ from the vortex line as 
\begin{equation}
v_s=\frac{\kappa\hbar}{2m_a\rho}.
\end{equation}
Here $\kappa$ is the strength of the vortex line.
This form of the velocity field implies the existence of a region close to the vortex axis
where the kinetic energy is large enough to break the Cooper
pairs. Hence the order parameter will be suppressed in the vortex
core and will heal to its bulk value over a length scale governed by the
coherence length $\xi_{\rm BCS}(T)=\hbar
v_F/\pi\Delta_{0}(T)$, with $v_F=\hbar k_F/m_a$ the Fermi 
velocity and $\Delta_{0}(T)$ the temperature dependent value of
the bulk gap away from the vortex core~\cite{VortexCoreNote}. 

Due to the single-valuedness of the order parameter the phase $\theta$
must return to the same value modulo $2\pi$ when going around
the vortex line. Hence the circulation
$\oint {\mathbf v}_s \cdot d{\mathbf{ \ell}}$ is restricted to integer
multiples of $h/2m_a$. In the present work we will concentrate
on vortices of unit circulation $\oint {\mathbf v}_s \cdot d{\mathbf{
\ell}}=h/2m_a$ .

In summary, a vortex line represents a topological defect in the
superfluid order parameter, around which the superfluid velocity field
${\mathbf v}_s$ is tangential. The quantization of the circulation
represents one the hallmarks of a superfluid, and therefore the
production and subsequent detection of quantized vortices in an
ultra-cold atomic Fermi gas would be a clear signature for
superfluidity in the system.

\section{Computational methods}
\label{numerics}

For a gas confined in a cylinder of radius $R$ and length $L$ it is
natural to work in cylindrical coordinates $(\rho,z,\varphi)$, where $\rho$
measures the perpendicular distance from the symmetry axis, $z$ is the
axial coordinate, and $\varphi$ is the azimuthal angle around
$\hat{z}$. In this coordinate system the order parameter can be written
as $\Delta({\mathbf r})=|\Delta(\rho,z)|\exp(-i\kappa\varphi)$, with $\kappa=0$ 
corresponding a phase with no vortex, and $\kappa=1$ for a singly
quantized vortex along the axis of symmetry. The mean-field density is
rotationally invariant: $n_{\sigma}({\mathbf r})=n_{\sigma}(\rho,z)$.

Assuming free motion along the cylinder axis, and imposing periodic
boundary conditions at $z=\pm L/2$, we write for the quasi-particle modes
\begin{eqnarray}
u_{\eta} ({\mathbf r}) & = & \rho^{-1/2} u_{nmk_z}(\rho) \,
\frac{e^{im\varphi}}{\sqrt{2\pi}}  \,
\frac{e^{ik_zz}}{\sqrt{L}} \nonumber \\
v_{\eta} ({\mathbf r}) & = & \rho^{-1/2} v_{nmk_z}(\rho) \,
\frac{e^{i(m+\kappa)\varphi}}{\sqrt{2\pi}}  \,
\frac{e^{ik_zz}}{\sqrt{L}}. 
\end{eqnarray}   
The allowed values of the angular momentum quantum number are 
$\{m=0,\pm1,\pm2,\ldots\}$, and $k_z=2\pi\ell/L$, with
$\{\ell=0,\pm1,\pm2,\ldots$\}. The radial functions
$(u_{nmk_z},v_{nmk_z})$ are taken to  be real. With these definitions the
BdG equations~(\ref{BdGeqs}) become
\begin{eqnarray}
\left[\begin{array}{cc}H_{m}&\Delta(\rho)\\
\Delta^*(\rho)&-H_{m+\kappa} 
\end{array}
\right]\left[\begin{array}{c}u_{nmk_z}(\rho)\\v_{nmk_z}(\rho)\end{array}
\right]=
E_{nmk_z}\left[\begin{array}{c}u_{nmk_z}(\rho)\\v_{nmk_z}(\rho)\end{array}
\right],
\label{BdGcyl}\label{BdGVortex}
\end{eqnarray}
where 
\begin{equation}
H_m=\frac{\hbar^2}{2m_a}\left[-\frac{\partial^2}{\partial \rho^2}
+\frac{(m^2-1/4)}{\rho^2}+k_z^2\right]+gn_{\sigma}(\rho)-\mu. 
\end{equation} 
These are the equations we solve self-consistently through an
iterative procedure.

By exploiting the symmetry of the BdG equations~(\ref{BdGeqs}), we can identify a {\emph{negative}} energy
solution with angular momentum $m$ with a {\emph{positive}} energy
solution with angular momentum $-m-\kappa$. We can therefore generate the
entire positive energy spectrum by solving Eq.(\ref{BdGcyl}) for $m\ge
0$ only, and using the transformation
\begin{equation}
E_{\eta}\rightarrow-E_{\eta}, \ \ \left( \begin{array}{c} u_{\eta} \\
v_{\eta} \end{array} \right) 
\rightarrow \left( \begin{array}{c} v_{\eta}^* \\ -u_{\eta}^*
\end{array} \right). 
\label{symmetry}
\end{equation}   
to find the eigenstates with $m<0$.

\subsection{Discrete Variable Representation}
The BdG equations in general must be solved numerically. Some of the
effects of the vortex that we are interested in, such as the
associated shifts in the critical temperature $T_c$ and in the  
ground state energy of the gas, are quite hard to calculate
numerically as they are very small compared with the corresponding
bulk values. For example, to obtain the vortex energy one needs to
subtract two large numbers (the ground state energy of the gas with
and without a vortex) to get a small number. This requires a very
accurate numerical scheme to solve the BdG eqns. Such a scheme is
provided by the Discrete Variable Representation (DVR) which recently
enabled the microscopic calculation of the vortex
energy~\cite{Nygaard2003}. DVRs are representations on a basis of
functions localized about discrete values of the coordinate. This
renders local functions of the coordinate operator approximately
diagonal within the DVR basis, making DVRs ideally suited for solving
self-consistent problems like the present one, where the matrix elements of
the pairing and Hartree fields (local functions) have to be evaluated at 
each iteration. In
addition the representation of the kinetic energy operator is
exact. The literature on DVRs is extensive and we shall only convey
the central points here. A detailed review of the framework can be
found in~\cite{Baye1986,Light2000}. 

A DVR exists when there is both a spectral basis of 
$N$ functions, $\phi_{n}(x)$, orthonormal over an interval [$a,b$] with
weight function $w(x)$ and a quadrature rule with $N$ points 
$x_{k}$ and weights $w_{k}$
\begin{equation}
\langle f|g \rangle \equiv \int_{a}^{b} dx \, w(x)f(x)g(x) \equiv
\sum_{k=1}^{N} w_kf(x_k)g(x_k). 
\label{intrule}
\end{equation}
This enables a set of coordinate eigenfunctions \mbox{$\lbrace \psi_i(x), 
i=1,N \rbrace $} to be defined with the property
\begin{equation}
\psi_i(x_k)=\delta_{ik}\sqrt{\frac{w(x_i)}{w_i}}.
\label{cond1}
\end{equation}
We expand the unknown functions $\psi_i(x)$ on the basis $\phi_n$
\begin{equation}
\psi_i(x)=\sum_{n=1}^{N} \phi_n(x)\langle\phi_n|\psi_i\rangle,
\end{equation}
and use the quadrature rule~(\ref{intrule}) and~(\ref{cond1}) to
evaluate the expansion 
coefficients. The coordinate eigenfunctions are then given by
\begin{equation}
\psi_i(x)=\sum_{n=1}^{N} \sqrt{w(x)w_i} \phi_n(x)\phi_n(x_i).
\label{uexp}
\end{equation} 
Since the $\psi_i(x)$ diagonalizes the coordinate operator, the matrix
element of any operator ${\mathcal{O}}(x)$, which is a local function
of $x$, is approximately diagonal within the DVR
\begin{equation}
\langle \psi_i|{\mathcal{O}}(x)|\psi_j\rangle \simeq {\mathcal{O}}(x_i)\delta_{ij},
\end{equation}
the approximation being due to the use of a truncated basis.
Furthermore, since the DVR involves an underlying spectral
representation, it is possible to evaluate matrix elements of parts of
the Hamiltonian exactly, if the $\phi_n(x)$ are chosen to be the
eigenfunctions of the corresponding operator (for example the kinetic
energy). 

For the problem of quantization in a cylinder the cylindrical
Bessel functions form an ideally suited basis for the DVR as suggested
in reference~\cite{Lemoine1994}. They are
orthogonal over the range $[0, R]$
\begin{equation}
\int_0^R d\rho \, \rho \, J_m(k_{im}\rho)
J_m(k_{jm}\rho)=\frac{\delta_{ij}}{w'_{im}}, 
\label{normcoor}
\end{equation}
where the coordinate normalization constant is given
by~\cite{Arfken1995} 
\begin{equation}
w'_{im}=\frac{2}{R^2 J^2_{m+1}(k_{im} R)}.
\end{equation}
Similarly, the Bessel functions are also orthogonal in momentum space:
\begin{equation}
\int_0^K dk \, k \, J_m(k\rho_{im})
J_m(k\rho_{jm})=\frac{\delta_{i j}}{{w}_{i m}},
\end{equation} 
with the momentum normalization 
\begin{equation}
{w}_{i m}=\frac{2}{K^2 J^2_{m+1}(K \rho_{im})}.
\end{equation}
The spatial and momentum grids are $\rho_{im}=z_{im}/K$,
and $k_{im}=z_{im}/R$, respectively, where 
$\{z_{im}, \ i=1,\ldots,N\}$ are the zeros of the Besselfunction of order $m$, defined through $J_m(z_{im})=0$. This
is a consequence of the boundary condition which states that the
wave function must vanish at $\rho=R$. 
Note that since $k_{Nm}=z_{Nm}/R=K_m$, and $\rho_N=z_{Nm}/K_m=R$, the
maximum momentum and the maximum value of $\rho$ are not independent,
but are inversely related to each other by the relation $R K_m =
z_{Nm}$. It was shown in~\cite{Lemoine1994} that a quadrature rule can
be associated with these grid points, provided weights are chosen to be
$w_{im}$ ($w_{im}'$) for integration over the spatial (momentum)
variable. In general there will be one spatial and one momentum grid
associated with each value of the angular momentum $m$.

With the Bessel function quadrature in place we can go ahead and
construct a DVR basis. As our orthonormal basis functions we choose
\begin{equation}
\phi_{i}(\rho) = \sqrt{w'_{im}} J_m(k_{im} \rho),
\end{equation}
where the $\sqrt{w'_{im}}$ is necessary to ensure that the basis set is
orthonormal, \emph{i.e.} $\langle \phi_k|\phi_l \rangle = \delta_{kl}$.
From (\ref{uexp}) we thus have for the coordinate eigenfunctions
\begin{equation}
\psi_{im}(\rho) = \sum_{n} \sqrt{\rho} \, \sqrt{w_{im}} w'_{nm} \,
J_m(k_{nm} \rho) J_m(k_{nm} \rho_{im}). 
\end{equation}
The radial functions $(u_{nmk_z},v_{nmk_z})$ can be expanded in terms of
the coordinate eigenfunctions, {\emph{i.e.}}
{\mbox{$u_{nmk_z}(\rho)=\sum_i \gamma_{im} 
\psi_{im}(\rho)$}}. The BdG equations will then be a set of
non-linear equations for the expansion coefficients
$\gamma_i$. Due to the properties of the coordinate eigenfunction the
value of the radial function on the grid points is simply
$u_{nmk_z}(\rho_{im})=\gamma_{im} \sqrt{\rho_{im}/w_{im}}$.  

We conclude this section with two important remarks. While the
transformation from the spectral basis to the coordinate
eigenfunctions is not strictly unitary, the numerical procedure is
nonetheless well defined, as the transformation can be made
unitary in the limit of large $N$~\cite{Lemoine1994}. Secondly,
although it appears that a separate grid is needed for each $m$ value,
we have found that in practice only two grids are needed,
one based on $J_0$ for even $m$ and one based on $J_1$ for odd $m$. Since
$u_{\eta}$ and $v_{\eta}$ for a vortex state correspond to
wavefunctions which differ by one unit of angular momentum, they will
be represented on different spatial grids. Fortunately, interpolation
is trivial in the DVR method. To interpolate from the $m=0$ to the
$m=1$ grid amounts to multiplying the vector of expansion coefficients
$\gamma_0$ with the transformation matrix given by
$B_{ij}=\psi_{i0}(\rho_{j1})$. The reverse transfomation is
$B_{ij}^{\dagger} = \psi_{i1}(\rho_{j0})$. For the purpose of solving the 
BdG equations the mean-fields are only represented on the odd $m$ grid.

\section{Thermodynamics}
\label{thermodynamics}

In this section, we present results for various thermodynamic
quantities of the vortex phase obtained by solving the BdG-eqns.\
numerically as described above. All calculations were done for a fixed
$N_{\sigma}=28000$. The radius an\ d length of the box were taken to be
$28.5 \ \mu{\mathrm{m}}$ and $11.4 \ \mu{\mathrm{m}}$,
respectively. For $^6$Li the scattering length is $-2160 \ a_0$, which
gives a bulk value of the transition temperature $T_{c0}=0.01
\ \mu{\mathrm{K}}$, and a Fermi temperature of
$T_F=0.70 \ \mu{\mathrm{K}}$ for the chosen density.   
 
\begin{figure}
\centering
\epsfig{file=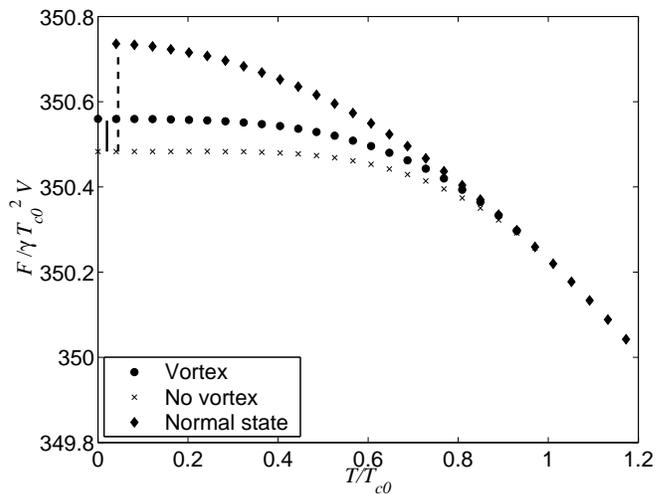,width=\columnwidth,angle=0}
\caption{Plot of the free energy per unit volume of $N_{\sigma}=28000$
fermions in the normal, and superfluid phase with and without a vortex
as a function of temperature. The solid and dashed vertical lines
represent analytic expressions for the vortex and condensation
energies, respectively.}  
\label{free_energy_fig}
\end{figure}

In figure~\ref{free_energy_fig} we plot the free energy
$\langle\hat{H}\rangle -TS$ as a function of the temperature $T$. 
The entropy is found as
\begin{eqnarray}
S &=& -k_{\rm{B}}\sum_{\eta} \, \left[ \, 
f(E_{\eta})\ln f(E_{\eta}) \right. \nonumber \\
\mbox{} && + \left. (1-f(E_{\eta}))\ln(1-f(E_{\eta})) 
\, \right],
\label{Entropy}
\end{eqnarray}
since the quasi-particles in our mean-field approach form an ensemble of non-interacting fermions~\cite{deGennes}.
We have calculated the free energy 
for the vortex phase, the superfluid phase without a vortex, and
for the normal phase. 
All have been normalized to $\gamma T_{c0}^2$, where
$\gamma=2\pi^2N(0)k_B^2/3$, and $N(0)=3n_\sigma/2\epsilon_F$ is the
density of states per unit volume (for a single component) at the
Fermi energy in the normal phase~\cite{Fetter1971a}. 
For $T=0$, the condensation energy density of
the superfluid without a vortex with respect to the normal phase is
$E_{\rm cond}/V=-N(0)\Delta^2_{0}/2$, with
$\Delta_0=8e^{-2}e^{-\pi/2k_F|a|}$ the bulk value 
of the superfluid gap. This condensation 
energy is indicated on the figure and we see that there is good
agreement with the numerical results. Furthermore, the vortex energy
per unit axial length for $T=0$ due to the loss of condensation energy
in the vortex core and the kinetic energy of the supercurrent around
the core 
is~\cite{Bruun2001a}   
\begin{equation}
\epsilon_v \approx \frac{\pi\hbar^2n_\sigma}{2m_a}\ln\left[D\frac{R}{\xi_{\rm
BCS}(0)}\right]. 
\end{equation}
The constant $D$ was determined numerically in ref.~\cite{Nygaard2003}
to be $D\approx 2.5$. This expression for the vortex energy is also
indicated on the figure.  
 
\begin{figure}
\centering
\epsfig{file=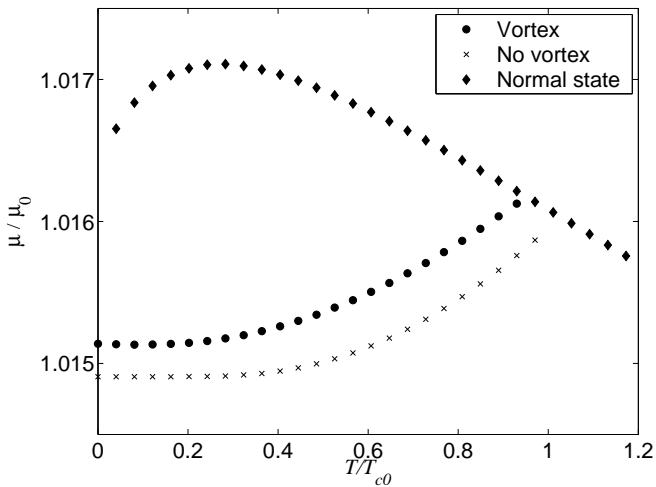,width=\columnwidth,angle=0}
\caption{Chemical potential in 
the normal, and superfluid phase with and without a vortex, as
determined by the constraint that $N_{\sigma}=28000$. We 
attribute the low temperature behavior of the normal phase chemical
potential to shell effects due to the finite volume~\cite{Schneider1998}.  
For the vortex state the transition temperature is shifted downwards.} 
\label{chemical_potential_fig}
\end{figure}

\begin{figure}
\centering
\epsfig{file=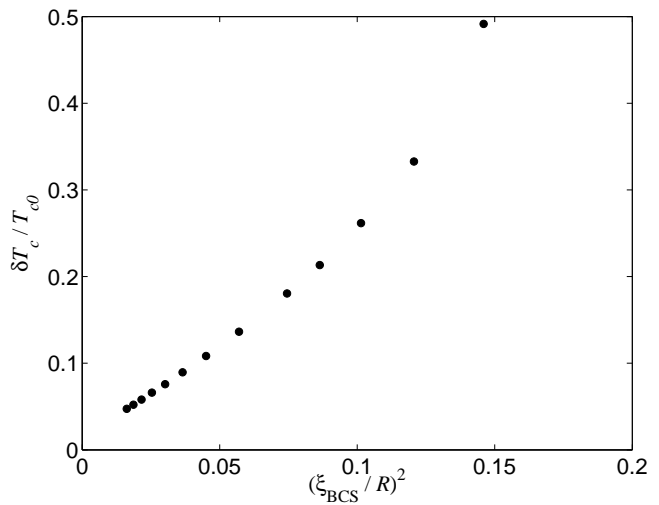,width=\columnwidth,angle=0}
\caption{The shift in the superfluid transition temperature for the vortex state relative 
to a bulk superfluid with no vortex, as a function of the radius of the confining cylinder at fixed density.}  
\label{Tc_shift_fig}
\end{figure}

As can be seen from Fig.~\ref{chemical_potential_fig}, the critical
temperature for the vortex phase $T_{cv}$ is  lower than that of the
bulk superfluid phase without a vortex $T_{c0}$. For the specific
parameters used, the difference is $1-T_{cv}/T_{c0}\approx 0.1$. This
difference can be understood as 
follows: The vortex phase becomes unstable with respect to the normal
phase when the extent of the vortex core becomes comparable to the
radius $R$ of the system. Since the size of the vortex is
${\mathcal{O}}(\xi_{\rm BCS})$, we can estimate $T_{cv}$ from the
condition $\xi_{\rm BCS}(T_{cv})\sim {\mathcal{O}}(R)$. Using  
$\Delta_{0}(T)\approx1.7\Delta_{0}(0)(1-T/T_{c0})^{1/2}$~\cite{deGennes}
for $0<1-T/T_{c0}\ll 1$, this yields  
\begin{equation}\label{Tcshift}
\frac{\delta T_c}{T_{c0}} \equiv 1-\frac{T_{cv}}{T_{c0}}\sim\frac{\xi_{\rm BCS}(0)^2}{R^2}\alpha_1
\end{equation}
where $\alpha_1$ is a number of order one. 
We now test this expression and determine the constant $\alpha_1$ by numerically
calculating the shift in the critical temperature $\delta T_c/ {T_{c0}}$ due to the presence of a vortex for various
radii of the system. The result is shown in Fig.~\ref{Tc_shift_fig}. We find 
that we get reasonable agreement with Eq.\
(\ref{Tcshift}) as $\xi_{\rm{BCS}}/R\rightarrow 0$ with a coefficient $\alpha_1\approx 2.3$. So
one can understand decrease in $T_c$ due to the presence of the  
vortex as a finite size effect which scales as $\xi_{\rm
BCS}(0)^2/R^2$. 

\begin{figure}
\centering
\epsfig{file=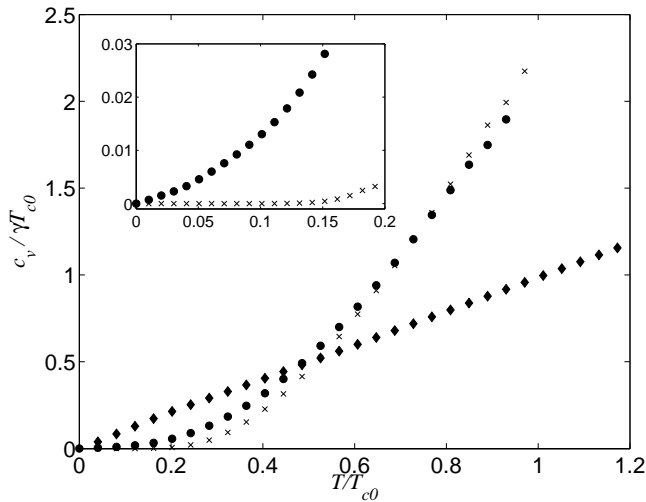,width=\columnwidth,angle=0}
\caption{Plot of the specific heat per unit volume in the normal, and superfluid phase
with and without a vortex. The inset shows the low temperature
behavior for the vortex state and the superfluid state without a
vortex (same symbols as in Figs.~\ref{free_energy_fig} and
\ref{chemical_potential_fig}).}  
\label{heat_capacity_fig}
\end{figure}

\begin{figure}
\centering
\epsfig{file=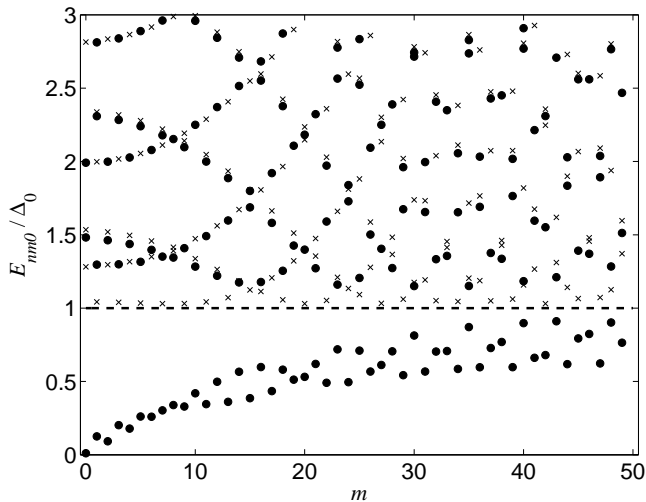,width=\columnwidth,angle=0}
\caption{Energy spectrum for the lowest quasi-particle states in a
superfluid with a vortex ($\bullet$) and the vortex free state
($\times$) at $T=0$. For clarity only the
energies of states with $k_z=0$ have been plotted. There are branches
of bound states for several values of $k_z$.}  
\label{qp_energies_fig}
\end{figure}

In Fig.~\ref{heat_capacity_fig} we plot the temperature dependence of
the heat capacity $c_{\rm V}=(T/V)\partial S/\partial T$ per unit
volume of the system. Again, we  
show for comparison results both for the system in the normal phase,
in the superfluid phase without a vortex, and in the vortex phase. For
a two component gas in the normal phase, we have $c_{\rm V\,
normal}=\gamma T$ for $T\rightarrow 0$. The heat capacity for the
superfluid phase without the vortex is exponentially damped by a factor
$\exp(-\beta\Delta_{0})$ for $T\ll T_c$ due to the gap in the
energy spectrum~\cite{Fetter1971a}.  
Figure~\ref{heat_capacity_fig} on the other hand shows that the heat
capacity in the vortex phase $c_{\rm V\, vortex}$ depends linearly on $T$
for low temperatures.  This linear $T$-dependence is due to the
presence of so-called core bound states in the vortex
phase. These are single-particle excitations which are  spatially
localized in the vortex core where the gap is small. The energy of
the core states is in general less than the bulk gap energy
$\Delta_{0}$ and they exist only for angular momentum quantum numbers
$m\ge0$~\cite{deGennes}. This corresponds to a quasi-particle
current around the vortex core in the opposite direction to that of the
vortex current. In a detailed analysis it was found that the energy
spectrum of the lowest bound core states with $0\le m\ll k_F\xi_{\rm
BCS}$ for $T=0$ is essentially gapless and given by 
\begin{equation}
E_{mk_z}\sim(m+1/2)\frac{\Delta_{0}^2}{\epsilon_F}\frac{h(\theta)}{\sin \theta}
\label{Vortexstates} 
\end{equation}
where $k_z=k_F\cos \theta$ and $h(\theta)$ is a function  of order
unity~\cite{Caroli1964}. In Fig.~\ref{qp_energies_fig}, we plot the
lowest quasi-particle energies as a function of $m$ for $k_z=0$
obtained from a numerical solution of 
Eq.~(\ref{BdGVortex}). The gapless branch associated with the core
states with energies less than $\Delta_{0}$ is clearly visible. The
$T=0$ density of vortex states per unit volume is calculated by
integrating Eq.\ (\ref{Vortexstates}) over $k_z$ which yields
\begin{equation}\label{Nvortex}
N_v(\epsilon)=N(0)\alpha_2\frac{\xi_{\rm BCS}^2}{R^2}
\end{equation}
for $0\ll\epsilon\ll\Delta_{0}$ where
$\alpha_2\sim{\mathcal{O}}(1)$~\cite{Fetter1969}. 
Thus, the density of bound core states per unit volume is the same, apart from 
a factor $\alpha_2$, as 
that of a cylindrical region of a single component gas in the normal
phase with radius $\xi_{\rm BCS}$ and length $L$. From this we
conclude that the low $T$ heat capacity per unit volume of the gas in
vortex phase $c_{\rm V\, vortex}$ associated with the core states is 
\begin{equation}
c_{\rm V\, vortex}\sim c_{\rm V\,
normal}\alpha_2\frac{\xi_{\rm BCS}^2}{R^2} 
\label{cVvortex}
\end{equation}
explaining the linear $T$-dependence of $c_{\rm V\, vortex}$ observed in 
Fig.~\ref{heat_capacity_fig}. A fit to the numerical data yields
$\alpha_2\approx 2$. We remark that a linear contribution to the heat
capacity has been observed for a superconductor in the mixed
state~\cite{Keesom1964}.

\section{Laser probing of the vortex phase}
\label{laserprobe}

Vortices are now routinely created in dilute BECs where they can be
detected by direct imaging of the cores, in which the density is
significantly suppressed. Unfortunately, such a procedure would be
very difficult to implement successfully for a dilute superfluid Fermi gas, where  there is 
no significant depletion of density in the vortex
core~\cite{Nygaard2003}.  
One way to observe the vortex is to measure the 
shift in the quadrupole mode frequencies which is directly proportional 
to the angular momentum per particle $\hbar/2$ associated with the 
supercurrent around the vortex core~\cite{Bruun2001a}.

In the present section, we investigate the feasibility of detecting the 
bound quasi-particle states in the vortex core through a
recently proposed laser probing
scheme~\cite{Torma2000,Bruun2001b}.
The laser probing scheme is similar to scanning tunneling microscopy
(STM) on a superconductor in that it relies on induced tunneling 
between a superfluid and a normal phase~\cite{Hess1989,Hess1990}. Whereas a STM probe uses a
bias voltage to transfer population across a superconducting-normal
interface existing between the normal microscope tip and the
superconducting substrate, the laser probe instead creates an
effective interface by coupling different internal states of the atoms
by laser fields. Specifically, a spin state $|\uparrow\rangle$,
which is Cooper paired with the state $|\downarrow\rangle$ is coupled
via laser field to a third state $|e\rangle$ that has been chosen such
that it does not participate in the pairing (either it does not have
strong attractive interactions with the two other states or the
disparity in chemical potentials is too large). Hence, the
$|e\rangle$ atoms define the normal part of the interface. If the
detuning of the laser from the atomic transition is
$\delta=\omega_{A}-\omega_L$, where $\omega_L$ is the 
laser frequency and $\omega_{A}$ the frequency splitting between
the level $|\uparrow\rangle$ and  $|e\rangle$, the rate of change in
the population of the $|e\rangle$ state (tunneling current)
$I=-\langle \dot{\hat{N}_e} \rangle$ is~\cite{Bruun2001b} 
\begin{eqnarray}
I(\delta) = &-& \frac{2\pi}{\hbar} \sum_{\eta,n} \left| \int d^3 r \,
\Omega({\mathbf r}) u_{\eta}({\mathbf r}) \Phi^*_{n}({\mathbf r})
\right|^2  
\nonumber \\
&\times& [f(E_{\eta})-f(E_{n})] \delta(E_{\eta}-\epsilon_{n}-\tilde{\delta})
\nonumber \\
&+&  \left|\int d^3 r \,
\Omega({\mathbf r}) v^*_{\eta}({\mathbf r}) \Phi^*_{n}({\mathbf r})
\right|^2 [1-f(E_{\eta})-f(\xi_{n})] 
\nonumber \\
&\times& \delta(E_{\eta}+\xi_{n}+\tilde{\delta}).\label{Current}
\end{eqnarray}
Here  $\tilde{\delta}=\mu_e-\mu+\delta\equiv \Delta\mu+\delta$ is the
effective detuning, $\mu_e$ the chemical potential of the $|e\rangle$
atoms, and  $\Phi_{n}$ their single particle wave functions with energy
$\xi_n$; $f(x)=[\exp(\beta x)+1]^{-1}$ is the Fermi function and
$\Omega({\mathbf r})$ the Rabi frequency. In the present analysis, we
assume for simplicity that the $|e\rangle$ atoms are non-interacting
such that their wavefunctions $\Phi_{n}$ are the eigenstates of the
confining cylindrical box. We consider the case of a constant laser profile
$\Omega({\mathbf r})=\Omega$. This gives the selection rule
${\mathbf{k}}_\uparrow={\mathbf{k}}_e$
where ${\mathbf{k}}_\uparrow$ is the momentum of an $|\uparrow\rangle$
atom coupled by the laser beam to an $|e\rangle$ atom with momentum
${\mathbf{k}}_e$. 

Let us now consider how the laser probing method can be used to probe
the presence of the core states. We examine two opposite cases of
interest: The case when there are initially no $|e\rangle$ atoms
present ($N_e=0$) and the case where there initially are an equal
number of $|\uparrow\rangle$ and $|e\rangle$ atoms present
($N_\uparrow=N_e$).  

From Eq.\ (\ref{Current}) it is straightforward to show that for the
total current we have $\int d \delta I(\delta)\propto N_e-N_\uparrow$. That 
is, the net current from the $|e\rangle$ atoms to the $|\uparrow\rangle$ atoms is 
proportional to the difference of initial  populations between the two
hyperfine states.  
Likewise, the total current from the core states trapped inside the
vortex is clearly proportional  
to the total number of core states $N_{\rm cs}$. Thus,
when there initially are no $|e\rangle$ atoms present ($N_e=0$) the
spectral weight of the current due the core states as compared to the
total current observed scales as $N_{\rm cs}/N_\uparrow$. Using $N_{\rm cs}\sim
N_v\Delta_{0}\pi R^2L$ with $N_v$ given by Eq.\ (\ref{Nvortex}), one
obtains that the current from the core states divided by the total
current scales as $\Delta_{0}\epsilon_F^{-1}\xi_{\rm BCS}^2R^{-2}\ll
1$. Thus, the signal from the core states is completely overwhelmed by
the bulk signal coming from the current out of the whole Fermi sea of
$|\uparrow\rangle$ atoms. We therefore conclude that it is most likely
not possible to probe the core states starting with initially no
$|e\rangle$ present. This conclusion is supported by numerical
simulations. 

\begin{figure}[!]
\centering
\epsfig{file=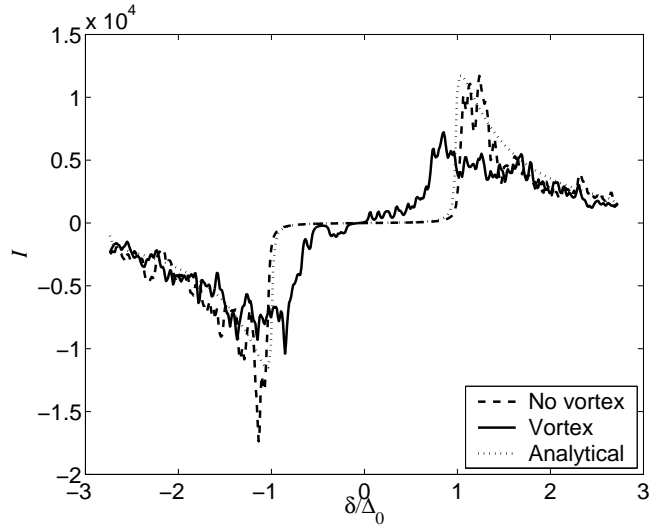,width=\columnwidth,angle=0}
\caption{The tunneling current as a function of detuning (in
units of the bulk value of the gap) for tunneling into filled state
from both a vortex state and a superfluid without a vortex. For
comparison Eq.~(\ref{anacurrent}) is also plotted. The profiles have
been shifted to compensate for the Hartree mean field shift $gn_{\sigma}$ of the
energies of the $|\uparrow\rangle$ atoms. If the $|\uparrow\rangle$
atoms are in the normal state no current flows due to Pauli blocking.}   
\label{laserprobe_fig}
\end{figure}

Let us therefore consider the case when there initially are an equal
number of $|\uparrow\rangle$ and $|e\rangle$ atoms present
($N_\uparrow=N_e$). In that way, the bulk signal of transitions of
$|\uparrow\rangle$ atoms deep within the Fermi sea is Pauli blocked
due to the presence of the $|e\rangle$ atoms since we have the selection rule 
${\mathbf{k}}_\uparrow={\mathbf{k}}_e$. One can then show from
Eq.\ (\ref{Current}) that the total signal scales as $\int d \delta
|I(\delta)|\propto N_\uparrow\Delta_{0}/\epsilon_F$, i.e.\ the current
is proportional to the total number of Cooper pairs. Thus, the bulk
signal is suppressed by a factor $\Delta_{0}/\epsilon_F$ compared to
the case when there are no $|e\rangle$ atoms present simply due to the Fermi 
blocking effect. The current due
to the vortex core states should therefore be easier to observe as it is not
overwhelmed by a huge background signal. In Fig.~\ref{laserprobe_fig}
we plot the $T=0$ laser probing current $I(\delta)$ for the case when
$N_e=N_\uparrow$. The effect of the Hartree field $gn_{\sigma}$ is
primarily to shift the entire profile to lower detunings $\delta$
since it shifts the  
energies of the $|\uparrow\rangle$ atoms by the amount $gn_{\sigma}$
whereas the $|e\rangle$ atoms are assumed non-interacting. In the
plot we have explicitly eliminated this overall shift for reasons of
clarity. We plot the current both when there is no vortex present and
when there is a vortex. In the case of no vortex present, the current
given by  Eq.\ (\ref{Current}) at zero temperature can be shown to be  
\begin{equation}
I=\pm\pi\Omega^2\rho(\delta) \, \Theta(\delta^2-\Delta_{0}^2)
\frac{\Delta_{0}^2}{\delta^2},  \label{anacurrent}
\end{equation}
where $\pm$ corresponds to $\delta>0$ and $\delta<0$ respectively and 
$\rho(\delta)=V(\Delta_{0}^2 /
\delta-\delta+2\mu)^{1/2}/2\pi^2$~\cite{Torma2000,Bruun2001b}. From Eq.\
(\ref{anacurrent}) it follows that there is no current for detunings
with $-\Delta_{0}<\delta<\Delta_{0}$. This can be interpreted as the   
laser signal has to provide a minimum energy $\Delta_{0}$ to break a
Cooper pair and generate a current. Equation (\ref{anacurrent}) is also shown on
Fig.~\ref{laserprobe_fig} and we see good agreement with the numerical
result when there is no vortex present. Note that since the
numerical calculations use a Lorentzian of width $\Gamma=0.01\Delta_0$
instead of $\delta(x)$ functions in Eq.\ (\ref{Current}), we have
convoluted Eq.\ (\ref{anacurrent}) accordingly. We see that the signal
when there is a vortex present is markedly different from the case
with no vortex. In particular, there is a significant current for
$|\delta|<\Delta_{0}$. This current is directly due to the presence of
the core states which have a pairing energy less
than $\Delta_{0}$. The signal from the vortex phase is finite for
$\delta\sim 0$ reflecting the fact that the energy spectrum of the
core states approximately given by Eq.\ (\ref{Vortexstates}) is
essentially gapless.  Thus, the existence of core states bound in the vortex is reflected 
in the current profile $I(\delta)$. 

\section{Conclusions}
We have studied the properties of a single vortex in a neutral superfluid with Fermi statistics 
using a microscopic weak coupling theory. The effect of the vortex on the free energy and the 
heat capacity of the system was examined and we provided various analytical expressions which agrees 
well with the numerical results. The vortex gives rise to the presence of core states bound in the 
vortex core. We examined  the spectrum of these states and also suggested a way to experimentally 
detect them. Apart from being of interest theoretically, it is not unlikely that our results will have 
experimental relevance in the near future due to the recent impressive experimental progress within the 
field of atomic Fermi gases.


\begin{thebibliography}{47}
\bibitem[{Vor()}]{VortexCoreNote}
\bibinfo{note}{We can estimate the radius of the cylinder in which interior
  superfluidity will be suppressed by equating the kinetic energy $\case{1}{2}
  m{\mathbf v}_s^2$ with the condensation energy per particle
  $3\Delta_0^2/4\mu$. This gives just $\sim\xi_{\mathrm{BCS}}$}.
\end{thebibliography}
\end{document}